\documentclass[sigconf]{acmart}

\AtBeginDocument{%
  \providecommand\BibTeX{{%
    \normalfont B\kern-0.5em{\scshape i\kern-0.25em b}\kern-0.8em\TeX}}}

\acmConference[MM '25]{Proceedings of the 33rd ACM International Conference on Multimedia}{October 27--31, 2025}{Dublin, Ireland}
\acmDOI{10.1145/3746027.3755266}
\acmISBN{979-8-4007-2035-2/2025/10}

\usepackage{multirow}
\usepackage{pifont}
\newcommand{\cmark}{\ding{51}}%
\newcommand{\xmark}{\ding{55}}%

\usepackage{bm}

\DeclareMathOperator*{\argmax}{argmax}
\usepackage{threeparttable}

\begin{document}

\title{MaXsive: High-Capacity and Robust Training-Free\\Generative Image Watermarking in Diffusion Models}


%
%
%
%
%
\author{Po-Yuan Mao}
\affiliation{%
  \institution{IIS, Academia Sinica}
  \country{Taiwan, ROC}}
%
\author{Cheng-Chang Tsai}
\affiliation{%
  \institution{IIS, Academia Sinica}
  \country{Taiwan, ROC}
  }

\author{Chun-Shien Lu}
\affiliation{%
  \institution{IIS, Academia Sinica}
  \country{Taiwan, ROC}
  }


\begin{abstract}
The great success of the diffusion model in image synthesis led to the release of gigantic commercial models, raising the issue of copyright protection and inappropriate content generation.
Training-free diffusion watermarking provides a low-cost solution for these issues. However, the prior works remain vulnerable to rotation, scaling, and translation (RST) attacks. Although some methods employ meticulously designed patterns to mitigate this issue, they often reduce watermark capacity, which can result in identity (ID) collusion. To address these problems, we propose MaXsive, a training-free diffusion model generative watermarking technique that has high capacity and robustness. MaXsive best utilizes the initial noise to watermark the diffusion model. Moreover, instead of using a meticulously repetitive ring pattern, we propose injecting the X-shape template to recover the RST distortions. This design significantly increases robustness without losing any capacity, making ID collusion less likely to happen. The effectiveness of MaXsive has been verified on two well-known watermarking benchmarks under the scenarios of verification and identification. 


\end{abstract}



\keywords{Attack, Diffusion Model, Generative Watermarking, Identification, Robustness}


\maketitle



\section{Introduction}
    Due to diffusion models' great success in generating high-quality images, well-trained commercial image synthesis models like Stable Diffusion (SD)~\cite{rombach2022high}, Muse AI, and Glide \cite{glide} were released to empower people to create high-quality images effortlessly.
    However, this brings up concerns about intellectual property protection. 
    Simultaneously, the introduction of AI security bills~\cite{EU_watermark,china_watermark,korea_watermark} highlights the urgent need of watermarking generated contents for protecting copyrights and tracing unauthorized use. 

To look back on the development of digital watermarking technologies~\cite{kutter1999towards,lin2001rotation,voloshynovskiy2001multibit,lu2006media}, they have already been recognized as an efficient mechanism for multimedia copyright protection in a post-processing manner in that the images are first generated and then watermarked.
Unlike traditional post-processing watermarking methods, diffusion generative watermarking integrates watermarking directly into the generation process. 
This mechanism of generative watermarking makes watermarking computationally efficient ({\em i.e.}, training-free), straightforward, and more secure~\cite{fernandez2023stable}. 
In particular, unlike learning model-based post-processing watermarking methods ({\em e.g.}, \cite{zhu2018hidden}), the inherent property of training-free in generative watermarking paradigm eliminates the need for additional training, further reducing computational overhead.

However, these algorithms remain vulnerable to rotation, scaling, and translation (RST) attacks. While Tree-Rings~\cite{wen2023treerings} and RingID~\cite{ci2024ringid} address rotation robustness using meticulously repetitive ring patterns, this design significantly reduces capacity, increasing the risk of ID collisions—where different watermark instances are mistakenly assigned the same identifier~\cite{xu2024invismark}. To address this trade-off, we investigate whether RST robustness can be achieved without relying on repetitive patterns, simultaneously resolving both RST distortions and ID collisions.

\begin{table*}[ht]
    \caption{Comparative analysis of in-process watermarking techniques for diffusion models. Note that RST resilience is assessed using Stirmark 3.1, where the distortion involves a combination of rotation, cropping, and resizing to keep image contents only.}\label{table1}
    \centering\begin{threeparttable}
    \begin{tabular}{c|c|c|c||c|c|c}
        \toprule
        \multirow{2}{*}{Methods} & \multicolumn{3}{c||}{Robustness} & \multicolumn{3}{c}{Capacity (Sec. \ref{sec:capacity_analysis})}\\
        \cmidrule{2-7}
        & Training&Noise layers & RST Resilience & $L$ & Ber / $\mathcal{N}$\tnote{b} & Capacity (bits)\\
        \midrule\midrule
        Stable Signature&\cmark&\cmark &\xmark & 48 & Ber & 48\\
        AquaLoRA&\cmark&\cmark &\xmark & 48 & Ber & 48 \\
        \midrule
         $\mbox{Tree-Rings}$&\xmark&\xmark  & $\Delta$\tnote{a} & 10 & $\mathcal{N}$ & 20.471 \\
         RingID&\xmark&\xmark &$\Delta$ & 11 & Ber&11 \\
        Gaussian Shading&\xmark&\xmark  &\xmark &256 & Ber & 256\\
        \midrule
        MaXsive&\xmark&\xmark & \cmark & \textbf{4,096} & $\mathcal{N}$& 8,384.9216\\
        \bottomrule
    \end{tabular}
    \begin{tablenotes}
            \item[a] $\Delta$ indicates the algorithm is able to address the rotation.
            \item[b] Ber represents that the algorithm uses a binary bit stream, while $\mathcal{N}$ denotes that the watermark is sampled from a normal distribution.
    \end{tablenotes}\end{threeparttable}
\end{table*}

In this paper, we propose MaXsive to bridge this gap. Unlike existing methods that depend on repetitive circular patterns to resist against rotations, MaXsive recovers rotations by an X-shaped template. Combined with non-discrete watermark design, MaXsive achieves a capacity of 8384 bits, far exceeding the capacities of previous approaches—$11$ bits for RingID \cite{ci2024ringid} and $256$ bits for Gaussian Shading \cite{yang2024gaussian}, making ID collisions highly unlikely. Furthermore, MaXsive surpasses all existing algorithms on the robustness benchmarks, Stirmark 3.1 \cite{petitcolas1998attacks,petitcolas2000watermarking} and WAVES \cite{an2024benchmarking}, in both verification and identification settings.\footnote{For verification, an embedded watermark is verified if it can be robust against image manipulations, while watermark identification means the ability to resolve ID collision.} Our contributions in this work are summarized as follows:
\begin{itemize}
\item High Capacity Training-free Algorithm: Based on Shannon entropy, MaXsive achieves significantly higher watermark capacity than previous training-free methods. This reduces the risk of ID collusion and enables real-world deployment without the need for additional fine-tuning.
\item Robust Diffusion Watermarking: MaXsive outperforms existing training-free diffusion-based generative watermarking algorithms, offering superior robustness in both identification and verification settings.
\item 
Novel Approach to RST Attack Resistance: MaXsive is the first to introduce the template for diffusion model watermarking, which effectively resolve RST (Rotation, Scaling, and Translation) attacks. Unlike previous algorithms using meticulously designed patterns, our template and watermark are not coupled together so as not to affect watermark capacity.

\end{itemize}

\section{Related Work}

\subsection{Non-Learning/Traditional Image Watermarking}
Digital watermarking plays a critical role in ownership protection and authentication, providing a secure method to track and validate digital assets.
However, a significant challenge is how to design watermarks that remain robustness against common image manipulations, such as compression and filtering, and geometric distortions like rotation and scaling.
In the literature, a broad range of studies have been proposed to address this issue.
For instance,~\cite{tang2003feature,dong2005digital} employed image normalization techniques to increase resilience against geometric transformations, while~\cite{pereira2000robust} used watermark embedding in the Fourier domain to strengthen robustness.
Additionally, other methods involve watermarking within geometry-invariant domains~\cite{5378607,kutter1999towards,647968,1227605}, feature-based watermarking~\cite{1532313,1036050,5378648,4454287,1673462,6486549}, and the use of periodic watermarks~\cite{kalker1999video,voloshynovskiy2001multibit}, each contributing to enhanced durability and reliability in digital watermarking applications.
\subsection{Watermarking Diffusion Models}
Diffusion model watermarking differs from traditional image watermarking by embedding the watermark directly during the image generation process.
In essence, images generated by diffusion models inherently contain watermarks.
There are two main approaches: (\emph{i}) fine-tuning-based watermarking~\cite{zhang2019robust,fernandez2023stable, feng2024aqualora,ma2024safe}, which utilizes the power of neural networks to learn and embed watermarks during training accompanying with the drawback of additional computational overhead and modification of model parameters, and (\emph{ii}) training-free watermarking~\cite{wen2023treerings,ci2024ringid,yang2024gaussian,arabi2025hidden}, which does not require retraining the model. The training-free methods, such as Tree-Rings~\cite{wen2023treerings}, embed watermarks into the initial noises in the Fourier domain, with further improvements by RingID \cite{ci2024ringid}, while Gaussian Shading~\cite{yang2024gaussian} cleverly projects the watermark onto the initial noise.
Although these training-free techniques show strong performance, their reliance on a limited number of keys makes them impractical for widespread, real-world use.


\section{Preliminaries}\label{sec:preliminary}
        In this section, preliminaries pertinent to the forthcoming introduction of the proposed method will be described.
         \begin{figure*}[t]
    \centering
    \includegraphics[width=0.8\linewidth]
    {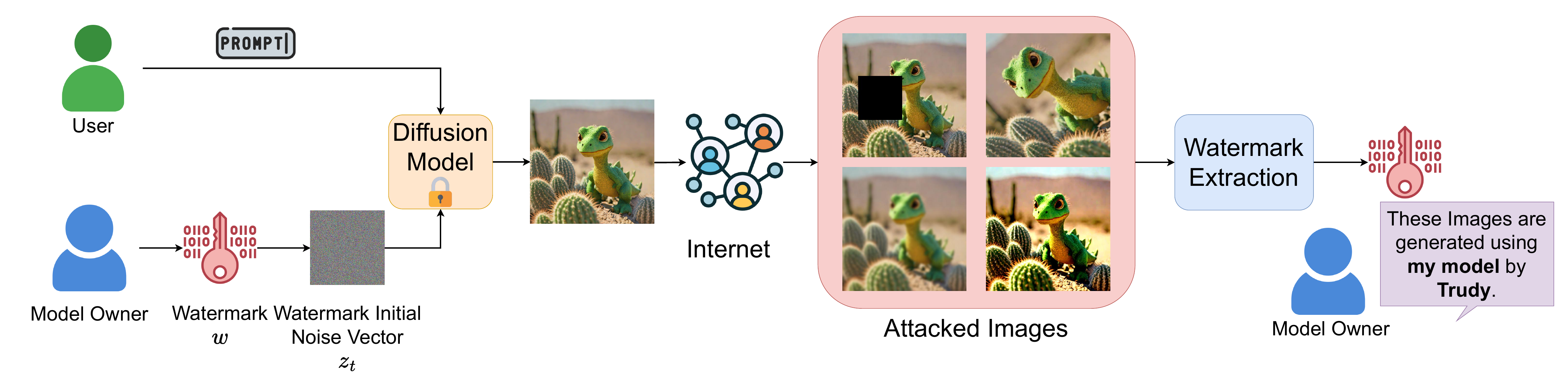}
    \caption{Real-World Applications of training-free diffusion watermarking algorithms.
    }
    \Description{\textcolor{red}{\bfseries Figure descriptions should not repeat the figure caption – their purpose is to capture important information that is not already provided in the caption or the main text of the paper.}}
    \label{real-world}
\end{figure*}
    \subsection{Latent Diffusion Models}\label{Sec: LDM}
        In the context of latent diffusion models (LDMs)~\cite{rombach2022high}, an initial noise $\bm{z}_{T}\in\mathbb{R}^{h\times w\times c}$, sampled from the standard Gaussian distribution ({\em i.e.}, $\bm{z}_{T}\sim\mathcal{N}(\bm{0},\mathbf{I})$), is used to generate an image in an RGB space.
        To generate an image, denoted by $\bm{x}\in\mathbb{R}^{H\times W\times 3}$, we first iteratively denoise $\bm{z}_{T}$ using the diffusion model $\epsilon_{\theta}$ to obtain its latent representation $\bm{z}_{0}$, and then use the decoder $\mathcal{D}$ to generate the image, \emph{i.e}., $\bm{x}=\mathcal{D}(\bm{z}_{0})$.
        The latent representation of $\bm{x}$ can be obtained using the encoder $\mathcal{E}$, \emph{i.e}., $\bm{z}_{0}=\mathcal{E}(\bm{x})$.

    \subsection{Denoising Diffusion Implicit Models}\label{sec:ddim}
        Here, we review the reverse and inverse processes of DDIM~\cite{song2021denoising} and introduce some notations.\footnote{In the literature on watermarking for diffusion models, the term ``inverse process'' was first introduced in Tree-Rings~\cite{wen2023treerings}, though the concept was initially mentioned in DDIM~\cite{song2021denoising} (where it was termed ``reversing the generation process'') and later described in~\cite{dhariwal2021diffusion} (as ``running the process in reverse'').}
        Given $\epsilon_{\theta}$ with $T$ timesteps, to obtain $\bm{z}_{0}$ using $\epsilon_{\theta}$, we iteratively apply $\epsilon_{\theta}$ to the latent samples $\bm{z}_{T},\bm{z}_{T-1},\dots ,\bm{z}_{t}, \dots,\bm{z}_{1}$.
        At $t$ step, we first estimate a predicted $\bm{z}_{0}$ as:
        \begin{equation}\label{eq:predict_0}
            \bm{z}^{t}_{0}=\frac{\bm{z}_{t}-\sqrt{1-\bar{\alpha}_{t}}\epsilon_{\theta}(\bm{z}_{t})}{\sqrt{\bar{\alpha}_{t}}},
        \end{equation}
        where $\bar{\alpha}_{t}=\prod^{t}_{i=0}(1-\beta_{t})$ and $\{\beta_{t}\}^{T}_{t=0}$ is a variance schedule, and then obtain $\bm{z}_{t-1}$ as:
        \begin{equation}\label{eq:next_step}
            \bm{z}_{t-1}=\sqrt{\bar{\alpha}_{t-1}}\bm{z}^{t}_{0}+\sqrt{1-\bar{\alpha}_{t-1}}\epsilon_{\theta}(\bm{z}_{t}).
        \end{equation}
        We denote the reverse process from $\bm{z}_{T}$ to $\bm{z}_{0}$ as $\mathcal{G}$, \emph{i.e}., $\bm{z}_{0}=\mathcal{G}(\bm{z}_{T})$.
        To obtain the initial noise of $\bm{z}_{0}$, at $t$ step, we follow:
        \begin{equation}\label{eq:inverse}
            \bm{z}_{t+1}=\sqrt{\bar{\alpha}_{t+1}}\bm{z}^{t}_{0}+\sqrt{1-\bar{\alpha}_{t+1}}\epsilon_{\theta}(\bm{z}_{t}).
        \end{equation}
        and denote the inverse process from $\bm{z}_{0}$ to $\bm{z}_{T}$ as $\mathcal{G}^{-1}$, \emph{i.e}., $\bm{z}_{T}=\mathcal{G}^{-1}(\bm{z}_{0})$.
        
    \subsection{Discrete Fourier Transform}\label{sec:dft}
    The discrete Fourier transform (DFT), a necessary technique for our proposed method, is reviewed here.
    Consider an image of size $N_{1}\times N_{2}$ as a real-valued function $i(p_{1},p_{2})$ defined on a grid $\{(p_{1},p_{2}) | p_{1}=0,1,\dots ,N_{1}-1,\ p_{2}=0,1,\dots ,N_{2}-1\}$.
    The DFT, denoted by $\mathcal{F}$, is defined as follows:
    \begin{equation}\label{eq:dft}
        I(k_{1},k_{2})=\sum^{N_{1}-1}_{p_{1}=0}\sum^{N_{2}-1}_{p_{2}=0}i(p_{1},p_{2})e^{-2\pi i\left( \frac{k_{1}}{N_{1}}p_{1}+\frac{k_{2}}{N_{2}}p_{2}\right)},
    \end{equation}
    and the inverse DFT (IDFT), denoted by $\mathcal{F}^{-1}$, is:
    \begin{equation}\label{eq:idft}
        i(p_{1},p_{2})=\frac{1}{N_{1}N_{2}}\sum^{N_{1}-1}_{k_{1}=0}\sum^{N_{2}-1}_{k_{2}=0}I(k_{1},k_{2})e^{2\pi i\left( \frac{p_{1}}{N_{1}}k_{1}+\frac{p_{2}}{N_{2}}k_{2}\right)}.
    \end{equation}

    \subsection{Problem Statement}
        We first introduce the scenario in which the proposed method can be applied, and then formulate the problem considered in this paper.

        \subsubsection{Application Scenarios}
            In a real-world scenario, as depicted in Figure~\ref{real-world}, watermarking involves the interaction between the model owner, users, and the internet.
            Unlike the scenario described in Stable Signature~\cite{fernandez2023stable}, where model providers ({\em i.e.}, model owners) deploy diffusion models directly to users, our scenario involves
            the model owner possessing a well-trained diffusion model and aiming to deploy it online, providing generation services via an application programming interface (API).
            To avoid the extra costs and quality degradation associated with fine-tuning the model to embed watermark information, the owner instead injects the watermark into non-training components.
            This watermark serves as proof of ownership and helps track who generates the images. However, since these images are shared online, users may modify them for their uses, thereby distorting the embedded watermarks accordingly.
            Therefore, the watermark extraction process must be robust to distortions.
            \subsubsection{Formulation}
            In this scenario, since the model owner does not want to pay the extra cost to train $\epsilon_{\theta}$ and the prompt $c$ is usually provided by users, a watermark $\bm{w}$ can only apply to $\bm{z}_{T}$, as described in Sec. \ref{sec:ddim}, where we can get the watermarked $\bm{z}^{\bm{w}}_{T}$ by $\mathcal{W}(\bm{z}_{T},\bm{w}) = \bm{z}^{\bm{w}}_{T}$ and $\mathcal{W}$ denotes the watermarking process. Consequently, the watermarked image $\bm{x}_{\bm{w}}$ is produced by:
            \begin{equation}\label{eq:sampling}
                \bm{x}_{\bm{w}} = \mathcal{D}(\mathcal{G}(\bm{z}^{\bm{w}}_{T})),
                \end{equation}
            where $\mathcal{D}$ is defined in Sec. \ref{Sec: LDM} and $\mathcal{G}$ is defined in Sec. \ref{sec:ddim}.
            
            During watermark extraction, the owner will get a distorted image $\bm{x}_{\bm{w}}'$. By applying the reverse function, the  watermark is extracted as:
            \begin{equation}
            \label{eq7}
                \hat{\bm{w}} =  \mathcal{W}^{-1} ( \mathcal{G}^{-1}(\mathcal{E} (\bm{x}_{\bm{w}}') ) ).
            \end{equation}
            Then, the distance between the original and extracted watermark patterns is measured. If the distance does not exceed a predefined threshold, the extracted watermark $\hat{\bm{w}}$ is judged to be the hidden one, leading to successful detection.

\begin{figure*}[ht]
    \centering
    \includegraphics[width=0.925\linewidth]{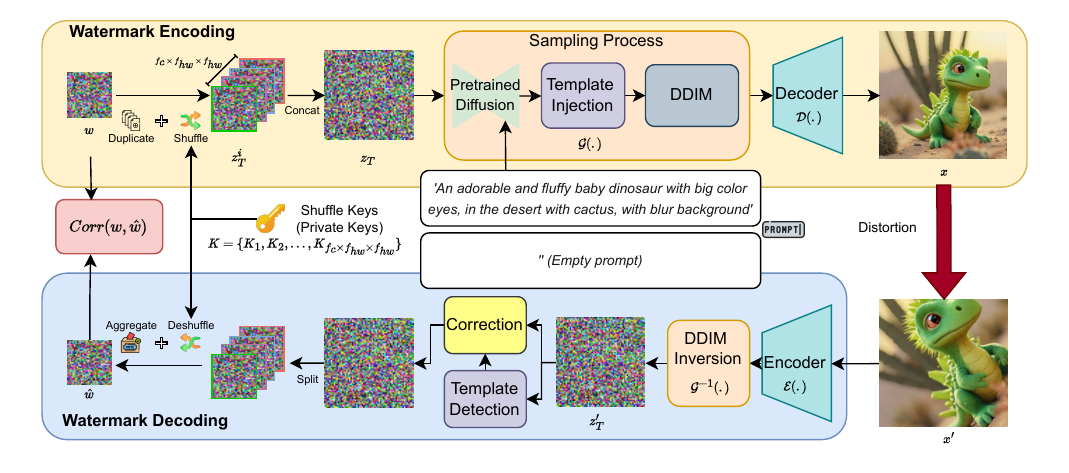}
    \caption{The framework of MaXsive. The watermark is a comparably small dimension vector sample from an ideal Gaussian distribution. The watermark is duplicated and encrypted by private keys, forming the input noise of the diffusion model.
    }
    \Description{\textcolor{red}{\bfseries Figure descriptions should not repeat the figure caption – their purpose is to capture important information that is not already provided in the caption or the main text of the paper.}}
    \label{fig:framework}
\end{figure*}

\section{MaXsive}
    In this section, we first outline the proposed method, MaXsive, in Sec.~\ref{sec:overview}, and then describe the watermark encoding and decoding processes in Sec.~\ref{sec:encoding} and Sec.~\ref{sec:decoding}, respectively. Finally, we provide capacity analysis in Sec. \ref{sec:capacity_analysis}.

    \subsection{Overview of MaXsive}\label{sec:overview}
        As depicted in Figure~\ref{fig:framework}, MaXsive consists of two processes: watermark encoding and watermark decoding. To restore from geometric distortions for robust watermark detection, we integrate an independent template estimate to enable recovery. Since this template does not directly affect the watermark, it avoids the capacity loss, a common phenomenon in Tree-Rings~\cite{wen2023treerings} and RingID~\cite{ci2024ringid}, which rely on repetitive ring watermark patterns.
        Additionally, our method avoids discretizing the watermark, thereby offering a larger capacity using the same number of elements. By integrating these design choices, our algorithm achieves both high capacity and strong robustness.
        
        In the encoding process, we generate a watermark by normalizing a vector sampled from the ideal normal distribution. This watermark is then duplicated, shuffled using a private key, and concatenated to match the input size of the diffusion model before being fed into it. During the sampling process (\emph{i.e}., Eq.~(\ref{eq:sampling})), we inject an invisible template at each timestep, ultimately obtaining the watermarked image $\bm{x}_{\bm{w}}$ by decoding the final latent representation.

        In the extraction stage, a potentially distorted image $\bm{x}_{\bm{w}}'$ is first encoded into its latent representation. This latent is then processed through two pathways: DDIM inversion and template detection. DDIM inversion reconstructs $\bm{z}_{T}$ from $\bm{z}_{0}$ by Eq.~(\ref{eq:inverse}), recovering the injected watermark, while template detection estimates the image's rotation angle in the Fourier domain. The recovered $\bm{z}_{T}'$ is then adjusted by the estimated angle. Finally, the extracted watermark $\bm{w}'$ is obtained by inverse shuffling and aggregation, with its similarity to $\bm{w}$ measured using the Pearson correlation coefficient. The details of these components are discussed in the following.

    \subsection{Watermark Encoding}\label{sec:encoding}
        MaXsive encodes generated images with a watermark.
        This process is based on the deterministic sampling process of DDIM~\cite{song2021denoising}, but with two modifications: (\emph{i}) Initial noise generation and (\emph{ii}) Template injection and design.

        \subsubsection{Initial Noise Generation}
        \label{sec:initial}
            The generation of a synthetic image using a diffusion model begins with sampling an initial noise from $\mathcal{N}(\bm{0},\mathbf{I})$.
            In contrast to Gaussian Shading~\cite{yang2024gaussian} and RingID~\cite{ci2024ringid}, which employ a binary bit stream as the watermark, our method aligns with Tree-Rings~\cite{wen2023treerings}, sampling the watermark $\bm{w}\in\mathbb{R}^{ \frac{h}{f_{hw}}\times \frac{w}{f_{hw}}\times \frac{c}{f_{c}}}$ from $\mathcal{N}(\bm{0}, \mathbf{I})$. This give us the advantage on the capacity (detailed analysis is in Sec.~\ref{sec:capacity_analysis}). To match the required input dimensions, $\bm{w}$ is replicated $f_c \times f_{hw} \times f_{hw}$ times, where $f_c$ and $f_{hw}$ denote the replication factors along different dimensions. However, when the length of $\bm{w}$ is not long enough, its sample means and variance may deviate from $0$ and $1$, potentially degrading image quality, as noted by~\cite{po2023synthetic}.
            To mitigate this, before duplication, we normalize the sampled values by normalizing their mean to 0 and standard deviation to 1.

            We find out that if we directly duplicate the watermark multiple times, the quality of generated watermarked images is still degraded because the resultant watermark contradicts the diffusion model assumption~\cite{nichol2021improved} (ablation in Sec. \ref{sec 6.3}). To introduce variability among duplications, each one is shuffled using a pseudo-random permutation determined by a shuffle key $K_i \in \bm{K} = \{K_1, K_2, \dots, K_{f_c \times f_{hw} \times f_{hw} } \}$. 
            This also ensures randomness in the initial noise, preventing predictable patterns. Eventually, these permuted elements $z_T^i$ are contacted, forming the initial noise $z_T$.

            
            
            \begin{figure}
                \centering
                \includegraphics[width=\linewidth]{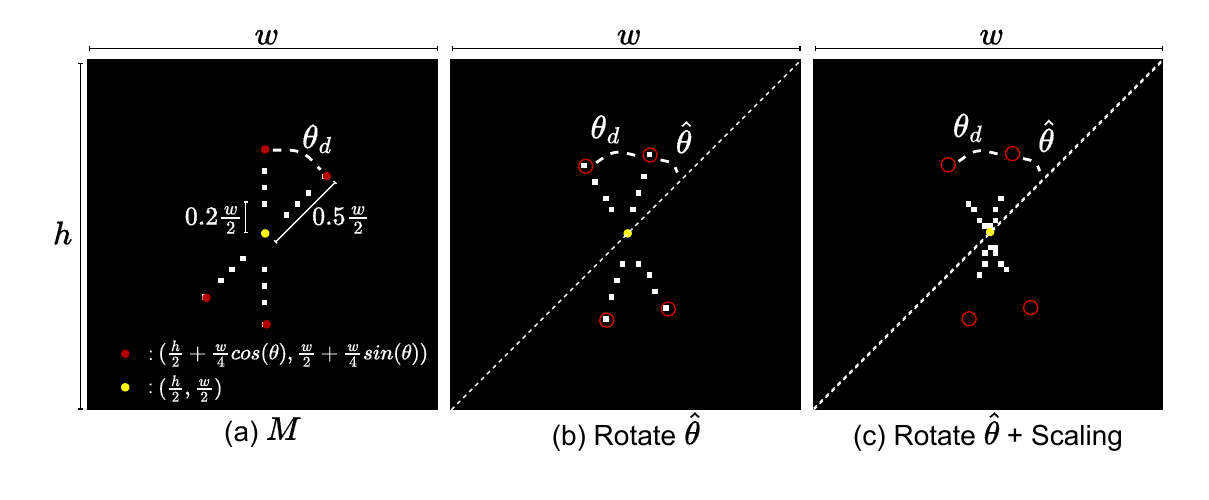}
                \caption{Illustration of Transformation: Rotation and Rotation with Scaling. (a) shows the defined template pattern. The yellow point indicates the center of the image, while the red points mark the outermost selected positions. (b) and (c) illustrate transformation behavior after applying rotation, and rotation combined with scaling, respectively. The red circle highlights a reference position used to verify whether the image has been scaled.}
                \label{fig:template}
            \end{figure}
        \subsubsection{Template Injection \& Design.}
            Although the initial noise generated by MaXsive is sufficient for watermark detection even under certain attacks (\emph{e.g}., JPEG compression), watermark detection significantly deteriorates under geometric distortions ({\em e.g.}, rotation, scaling, and translation (RST) attacks). To counter such attacks, we use an X-shaped template to uncover attack effects, which is the main challenge for diffusion watermarking algorithms. In contrast to the meticulously designed watermark pattern used in Tree-Rings~\cite{wen2023treerings} and RingID~\cite{ci2024ringid}, our template injection and watermark embedding are not coupled together so as not to affect watermark capacity.

            Two challenges are required to designed the X-shaped template: (\emph{i}) Template injection cannot degrade the quality of the generated images and (\emph{ii}) The template must be robust to various distortions. In order to address the two issues, we need to answer (\emph{i}) where to inject and (\emph{ii}) how to inject by describing template pattern design and template injection below.

            \paragraph{Template Pattern Design.}
             We employ an X-shaped binary mask $\bm{M}$ with the dimension of $h\times w$ to define the precise region for template injection, as illustrated in Figure~\ref{fig:template}, where $h=w$ in this paper. 
             The binary mask is structured such that ``1's'' in the mask indicate the selected positions, where the template is to be injected. These positions are distributed uniformly along two intersecting lines that form the X-shape. The intersection of the lines occurs at the center of the image, specifically at coordinates ($\frac{h}{2}$, $\frac{w}{2}$), which corresponds to the midpoint of the $z_T$. The two lines are oriented at an angular difference of $\theta_d$ degrees relative to each other.
              To yield a good compromise for reliable detection and minimal alteration to the image structure, a line is defined to be composed of 8 points, located in the range of $0.2\frac{w}{2}$ to $0.5\frac{w}{2}$ with an interval of $0.1\frac{w}{2}$ between two points. 

\begin{figure}[ht]
    \centering
    \includegraphics[width=\linewidth]{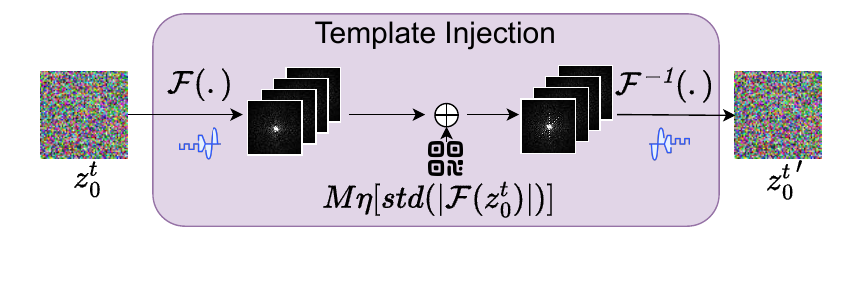}
    \caption{Template injection workflow. $\mathcal{F}$ and $\mathcal{F}^{-1}$ are defined in Eq.~(\ref{eq:dft}) and Eq.~(\ref{eq:idft}), respectively.
    }
    \Description{\bfseries Figure descriptions should not repeat the figure caption – their purpose is to capture important information that is not already provided in the caption or the main text of the paper.}
    \label{template injection}
\end{figure}

            \paragraph{Template Injection.}
            Unlike post-processing-based watermarking methods~\cite{pereira2000robust,kang2003dwt}, where the watermark is added after generation, our approach integrates the template injection directly into the sampling process. Directly injecting the template into $z_T$ during sampling, leads to interference, making it unstable for detection. Hence, we propose injecting the template by guiding the sampling process. More specifically, we notice that the second term of Eq.~(\ref{eq:next_step}) is the direction pointing to the origin noise $\bm{z}_t$ and
            $\bm{z}^{t}_{0}$ in Eq.~(\ref{eq:predict_0}) is the direction to predicted $\bm{z}_0$. To guide $\bm{z}_{t-1}$ towards the template direction, we replace $\bm{z}^{t}_{0}$ with $\bm{z}^{t}_{0} +\bm{M}\eta$, where $\eta$ is the parameters controls strength of the template. However, directly modifying $\bm{z}_0^{t}$ leads to a decrease in quality of the final images. To address this, we inject the template into the Fourier domain, {\em i.e.}, $\mathcal{F}(\bm{z}^{t}_{0} )$. The whole template injection process is defined as:
            \begin{equation} 
            \label{eq9}
            {\bm{z}_0^{t}}' = \mathcal{F}^{-1} (\mathcal{F}(\bm{z}^{t}_{0} )+ M\eta [  std( |\mathcal{F}(z_0^t )|)]),
            \end{equation}
            where ``$std$'' denotes the standard deviation.
            The template should balance the trade-off between visibility and reliable detection. However, since the injection process occurs at every sampling step—with each step exhibiting different distributions—a fixed template fails to meet this requirement effectively. To address this, we use $std(|\mathcal{F}(z_0^t)|)$ as a reference to adaptively adjust the injection magnitude. The overall injection process is illustrated in Figure~\ref{template injection}.

\subsection{Watermark Decoding}\label{sec:decoding}
        MaXsive decodes a possibly watermarked image $x'$ to recover the watermark, as defined in Eq.~(\ref{eq7}).
        Specifically, watermark decoding is based on the output $z'_T =\mathcal{G}^{-1}(\mathcal{E}{(x')})$ of the DDIM's inverse process, with three additional steps: (\emph{i}) Detection of the template,  (\emph{ii}) Correction of the initial noise, and (\emph{iii}) Estimation of the watermark.

        \subsubsection{Detection of Template.}
        \label{Detection of Template}
            We address the problem of template detection by framing it as a maximum likelihood estimation problem. Given $\mathcal{F}(\bm{z}'_{T})$ and the prior knowledge of template's shape, we aim to detect the lines that cross at the center. Specifically, given $z_{T}$ of size $h\times w$, as indicated in Sec.~\ref{Sec: LDM}, we
            set the origin point at ($\frac{h}{2}$, $\frac{w}{2}$).
            Given a line passing through the origin point with degree $\theta$, the set of points on this line can be represented as:
            \begin{equation}\label{eq:line_representation}
                L_{\theta}\coloneq\left\{ (p_{1},p_{2})\left| \ p_{2}-\frac{w}{2}=\tan\left(\theta\right)\left( p_{1}-\frac{h}{2}\right)\right\}\right. .
            \end{equation}
            Since the template is injected by adding $\eta [std ( |\mathcal{F} ( \bm{z}^{t}_{0} ) | ) ]$ to specific positions of all $\mathcal{F} ( \bm{z}^{t}_{0} )$ for $0\leq t\leq T$, as indicated in Eq.~(\ref{eq9}), the magnitudes located at $\bm{M}$ are local extrema. 
            Therefore,
            we perform a greedy algorithm
            to find the angle $\hat{\theta}$ that maximizes the average magnitude belonging to $L_{\hat{\theta}}$.
            Specifically, we calculate
            the mean of the magnitude of every candidate ranging from 0 to 360 degrees
            and formulate the objective function
            as follows:
            \begin{equation}
            \label{eq:theta_hat}
                \hat{\theta}=\argmax_{\theta}\frac{1}{n} \sum_{(p_{1},p_{2}) \in L_{\theta}\cup L_{\theta + \theta_d}} \left|\mathcal{F}(\bm{z}'_{T})(p_{1},p_{2})\right|,
            \end{equation}
            where $n$ is the number of pixels belonging to the X-shape template.
            
            
            The usage of the template is not only for the detection of the degrees of rotation but also possesses the benefit of detecting whether the image has been scaled, which is a common geometric distortion.\footnote{There are two common scaling effects in displaying images. One is to preserve all the image content without cropping and another one involves cropping.} However, the scaling plus cropping attack will cause a huge impact on Tree-Rings~\cite{wen2023treerings} and RingID~\cite{ci2024ringid}. Their meticulously designed watermark pattern cannot survive after this kind of scaling. In contrast, our X-shaped template can be used to resist against this kind of scaling since there is a corresponding geometric transformation on the template~\cite{pereira2000robust}. As illustrated in the third column of Figure~\ref{fig:template}, the X-shaped template exhibits a consistent transformation behavior: it becomes increasingly concentrated toward the center when the image is scaled up. We leverage this property to enhance robustness against geometric attacks.
            
            Specifically, as the rotation angle has been detected by Eq.~(\ref{eq:theta_hat}), we can verify whether the template is still in these positions (indicated by the red circles in Figure~\ref{fig:template}) by examining their magnitudes. If the magnitudes of the designated positions at the template exceed those of their neighboring positions, we consider the image is not rescaled. However, due to the extremely low resolution of $\mathcal{F}(\bm{z}'_{T})$, the outermost positions may not align precisely with the exact angle. To address this, we also check adjacent angular positions to improve detection robustness. Eventually, this will serve as extra information during the correction.
\begin{table*}[!t]
\centering
\caption{Verification via WAVES.}
\label{main_table}
\resizebox{0.95\textwidth}{!}{
    \begin{tabular}{r|  c c| c c c c c c | c }
    \toprule
    \multirow{ 2}{*}{Methods} & \multicolumn{2}{c|}{Quality}&\multirow{ 2}{*}{Clean}  &\multicolumn{5}{c|}{Attacks in WAVES} &\multirow{ 2}{*}{Avg} \\
    & CLIP $(\uparrow)$ & FID $(\downarrow)$&  & Geometric & Photometric & Degradation & Adversarial & Regeneration\\
    \midrule
    Tree-Rings &  32.43 & 17.32 & 0.71 &  0.21 & 0.51 & 0.26 & 0.39 & 0.07 & 0.29\\
    RingID   & 31.56 & 26.30 & 1.00  & 0.71 & 1.00 & 0.95 &1.00& 1.00 &0.93 \\
    Gaussian Shading  &  32.20 & 17.70& 1.00 &  0.58 & 1.00 & \textbf{0.97} &1.00& 1.00 & 0.91\\
    \midrule
    Ours ($\eta = 5$)&  32.35 & 17.35 & 1.00 &  \textbf{0.73} &  \textbf{1.00} & 0.95 & \textbf{1.00} &\textbf{1.00}&\textbf{0.94}\\
    \bottomrule
\end{tabular}
}
\end{table*}
\begin{table}
    \caption{Verification via Stirmark.}\label{stirmark}
    \begin{tabular}{r|c  c }
    
    \toprule
         Methods & Stirmark All & Stirmark RST  \\
         \midrule
         Tree-Rings & 0.24 & 0.01 \\
        RingID &  0.85 &  0.34\\
        Gaussian Shading & 0.86  & 0.27 \\
        Ours ($\eta = 5$) & \textbf{0.87} & \textbf{0.70}\\
        
        \bottomrule
    \end{tabular}
\end{table}
        \subsubsection{Correction of Initial Noise.}\label{sec:correction_initial_noise}
            Restoration has proven effective in mitigating geometric distortions within the domain where the distortion was applied \cite{awrangjeb2006global}. However, applying restoration directly to the image $x$ requires a second DDIM inversion, which is the most time-consuming step. To avoid performing inversion twice, we empirically find that restoration on $z'_T$ is also effective. Specifically,
            knowing that the image has been rotated by $\hat{\theta}$ degrees and whether it has been scaled,
            we can start restoration. We begin by restoring the rotation in that we rotate
            $z'_T$ counterclockwise with the detected angle $\hat{\theta}$ about the center of the image located at ($\frac{h}{2}$,$\frac{w}{2}$). Then, we rescale the $z'_T$ to $\frac{h}{\gamma}\times\frac{w}{\gamma}$, where $\gamma$ is the scaling parameter which is calculated by $sin(\hat{\theta}) + cos(\hat{\theta})$. (The derivation of $\gamma$ is shown in Sec. \ref{sec:gamma} of Appendix). Finally, we adjust the dimensions of the corrected $z'_T$ using zero-padding to restore the original size.
            
            


        \subsubsection{Estimation of Watermark.}
        To extract the embedded watermark, we start by uniformly slicing the latent representation $\bm{z}'_T$ into segments that match the dimensions of the original watermark. Each of these segments is then reordered using the private key $\bm{K}$, which is discussed in Sec.~\ref{sec:initial}, to reverse the shuffling operation applied during the embedding phase. After deshuffling, we perform an aggregation step by computing the average of the deshuffled segments. Subsequently, through empirical evaluation, we observed that using the Pearson correlation as the distance function yields better performance for similarity measurement in this context, outperforming L1-norm adopted in previous methods such as Tree-Rings and RingID. This suggests that Pearson correlation is more robust in capturing the structural similarities between the extracted and original watermarks under our proposed framework. 

    \subsection{Watermark Capacity Analysis}\label{sec:capacity_analysis}
        In this section, we introduce the framework to quantify the capacity of watermarks whose elements are sampled from the standard Gaussian distribution, in comparison with those sampled from the Bernoulli distribution with a parameter of $0.5$, denoted by $\mbox{Ber}(0.5)$.

        For a fair comparison, we consider each watermark $\bm{w}$ as a vector of random variables $[w_{1}, w_{2}, \dots , w_{L}]$, where $L$ is the number of elements in $\bm{w}$ and $w_{1}, w_{2}, \dots , w_{L}$ are independent and identically distributed (IID).
        Therefore, the values of $L$ for the methods discussed in this paper can be easily determined, as shown in Table~\ref{table1}.
        However, the random variables of different methods may follow different distributions.
        For example, the random variables of Stable Signature~\cite{fernandez2023stable}, AquaLoRA~\cite{feng2024aqualora}, and Gaussian Shading~\cite{yang2024gaussian} follow $\mbox{Ber}(0.5)$, whereas those of Tree-Rings~\cite{wen2023treerings}, RingID~\cite{ci2024ringid}, and MaXsive follow $\mathcal{N}(0,1)$.
        As a result, we need to quantify the random variables that follow different distributions on a fair comparison basis.
        To achieve this, we introduce the Shannon entropy (hereafter referred to as entropy), which is defined as:
        \begin{equation}\label{eq:shannon_entropy}
            H(X)=\mathbb{E}[-\log_{2}{p(X)}],
        \end{equation}
        where $X$ is a random variable that follows probability distribution $p$, $\mathbb{E}$ denotes the expectation, and $\log$ denotes the logarithm to the base 2.
        Using entropy (\emph{i.e}., Eq.~(\ref{eq:shannon_entropy})), we can quantify the random variables of the Bernoulli distribution and the standard Gaussian distribution for a more comprehensive comparison of the watermark capacity among different methods.
        
        Specifically, for the Bernoulli distribution, its entropy is
        \begin{equation}\label{eq:bernoulli}
            H_{b}=-\left(\frac{1}{2}\log_{2}\frac{1}{2}+\left(1-\frac{1}{2}\right)\log_{2}\left(1-\frac{1}{2}\right)\right)=1.
        \end{equation}
        On the other hand, the entropy of the standard Gaussian distribution is
        \begin{equation}\label{eq:gaussian}
            H_{g}=-\mathbb{E}\left[\log_{2}\left(\frac{1}{\sqrt{2\pi}}e^{-\frac{x^{2}}{2}}\right)\right]=\frac{1}{2}\log_{2}\left(2\pi{e}\right)\approx 2.0471.
        \end{equation}
        Since the random variables of different distributions are quantified, the watermark capacity of each method can be determined and compared.
        Thus, the watermark capacity is computed by
        \begin{equation}\label{eq:watermark_capacity}
            C = \left\{
            \begin{array}{l}
                L\times H_{b}\quad \mbox{if }w_{1}, w_{2}, \dots ,w_{L}\sim\mbox{Ber}(0.5)\\
                L\times H_{g}\quad \mbox{if }w_{1}, w_{2}, \dots ,w_{L}\sim\mathcal{N}(0,1)
            \end{array}
            \right.,
        \end{equation}
        where $H_{b}$ and $H_{g}$ are the computational results from Eq.~(\ref{eq:bernoulli}) and Eq.~(\ref{eq:gaussian}), respectively.
        Finally, we compute the watermark capacity of each method using Eq.~(\ref{eq:watermark_capacity}) and present the results in Table~\ref{table1}.

\section{Experimental Results}
\subsection{Setup}

\begin{table*}[!t]
\centering
\caption{Identification via WAVES. The performance result for each distortion is presented in Table~\ref{tab:WAVES_all_id} of Appendix.}
\label{identification}
    \begin{tabular}{r|   c  c c c c c  |c  }
    \toprule
    \multirow{ 2}{*}{Methods}&\multirow{ 2}{*}{Clean} & \multicolumn{5}{c|}{Attacks in WAVES} &\multirow{ 2}{*}{Avg}\\
     & &Geometric & Photometric & Degradation & Adversarial & Regeneration &\\
    \midrule
    \midrule
    Tree-Rings & 0.11 & 0.01 & 0.04  & 0.01  &  0.04  & 0.01  & 0.02  \\
    RingID   & 0.42 & 0.27 & 0.42 & 0.33&  0.42 & 0.40 & 0.37 \\
    Gaussian Shading  & 1.00 & 0.36 & 1.00 &  0.72 &  1.00 & 0.99& 0.81 \\
    \midrule
    Ours ($\eta = 5$) & \textbf{1.00} & \textbf{0.82} & \textbf{1.00} & \textbf{0.93} &  \textbf{1.00} &\textbf{1.00} & \textbf{0.95}\\
    \bottomrule
\end{tabular}
\end{table*}

In all our experiments, we compared the performance of MaXsive with three training-free diffusion watermarking algorithms with their optimal setting, including Tree-Rings \cite{wen2023treerings}, RingID \cite{ci2024ringid}, and Gaussian Shading \cite{yang2024gaussian} with their original settings. In detail, we used stable diffusion 2.1 \cite{rombach2022high} to generate watermarked images by the prompt ``A photo of a [class],'' where [class] is the ImageNet label. During the inversion diffusion process, we followed the settings of previous work \cite{wen2023treerings,ci2024ringid,yang2024gaussian}, using the DPM solver \cite{lu2022dpm} to revert images to their initial noise state with a blank prompt. We report a true positive rate (TPR) corresponding to a $1e-3$\footnote{In Tree-Rings and RingID, a true positive rate (TPR) corresponding to FPR at $1e-2$ was evaluated on stable diffusion prompts \cite{SD_dataset}.} false positive rate (FPR) to evaluate the watermark detection performance. Additionally, we evaluated watermarked image quality using CLIP-Score and Fréchet Inception Distance (FID) for 10,000 images.

Regarding verification, we evaluated 1000 images for each algorithms with the single distortion, regeneration, and adversarial attacks based on the WAVES benchmark \cite{an2024benchmarking}. For single distortion, we considered five evenly spaced distortion strengths between the defined minimum and maximum values. Regeneration was performed using the highest strength—7 for the VAE and 200 noise/de-noising timesteps for Stable Diffusion v1.4. Adversarial attacks were applied with a perturbation level of 8/255.
Furthermore, we conducted evaluations using Stirmark 3.1 \cite{petitcolas1998attacks,petitcolas2000watermarking}, a traditional but popular benchmark providing a comprehensive assessment of robustness against geometric distortions and signal processing manipulations.
Actually, an image undergone Stirmark will generate $88$ Stirmark attacked images. In the following, we will denote ``Stirmark All'' to represent all Stirmark attacks and ``Stirmark RST'' to indicate the attacks involving rotation, scaling, and translation.

For identification, we considered a scenario with 4,096 users, each generating 5 images. We evaluated on a single scale for each attack in WAVES, with distortion strengths set to: JPEG 10, rotation 45°, 87.5\% random C$\&$S, blurring kernel 15, noise std 0.1, brightness 2, and contrast 2. Regeneration was performed using the highest strength, {\em i.e.}, 7 for the VAE and 200 timesteps for noise adding/de-noising, in Stable Diffusion v1.4. Adversarial attacks were applied with a perturbation level of 8/255.

\subsection{Verification Results}


The verification results obtained under WAVES are shown in Table \ref{main_table}. To further validate the effectiveness of our method in addressing geometric distortions, comparison conducted under Stirmark 3.1 is shown in Table \ref{stirmark}. 
The use of $\eta=5$ in Eq. (\ref{eq9}) for template injection in MaXsive is to balance between image quality (see Table \ref{abalation:template}) and robustness.
Since the effect of rotations in Stirmark involves cropping and resizing, resulting in a combination of more than one geometric distortions. This causes a misalignment of the RingID and Tree-Rings patterns, which leads to a decline in robustness performance. However, MaXsive leverage the advantage of the template, designed to mitigate the impact of these distortions. We provide the detailed performance results for each distortion of Stirmark and WAVES in Table~\ref{tab:stirmark_all} and Table~\ref{tab:WAVES_all}, respectively, in Sec. \ref{Appendix: Details} of Appendix.

\subsection{Identification Results}

The advantage of high-capacity watermarking becomes especially prominent in identification tasks, where the objective is to accurately determine the identity of the user who created a given image. As shown in Table~\ref{identification}, this task places significant demands on the watermarking system's ability to embed unique and robust user information.

Tree-Rings and RingID struggle in this setting due to limited embedding capacity, which is insufficient to represent all 4,096 users. This limitation leads to frequent ID collisions, thereby severely degrading identification performance. Regarding Gaussian Shading, it achieves notably better results in the identification setting, owing to its ability to generate a sufficiently large number of unique keys. However, it is significantly vulnerable to geometric distortions. In contrast, MaXsive clearly outperforms these methods in the identification task.






\section{Ablation Study}
\subsection{Effect of Shuffler}

\label{sec 6.3}
\begin{figure}[h]
\centering  
\includegraphics[width=0.9\columnwidth]{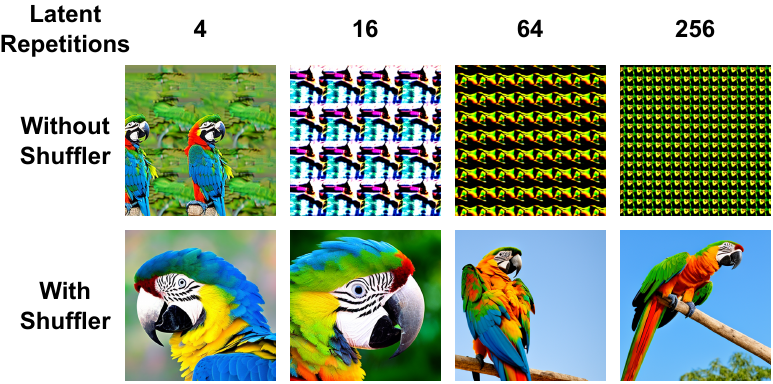}
\caption{Comparison of with and without shuffler in different repetition times.
}
\label{compare_repeat}
\end{figure}

The shuffler acts as a crucial component to preserve image quality by maintaining the condition necessary to meet the assumption of diffusion models in that the initial noise follows a Gaussian distribution. 
However, as described in Sec.~\ref{sec:encoding} and Figure~\ref{fig:framework}, when the noise used in the diffusion model is obtained from concatenating duplicated patterns, this assumption is violated. As shown in the top row of Figure~\ref{compare_repeat}, repetition introduces non-Gaussian artifacts that can destabilize the model. 
In our method, by shuffling the input noise, this randomness helps ensure that the noise distribution remains close to the ideal Gaussian, as shown in the bottom row of Figure~\ref{compare_repeat}.
\subsection{Distortion effect on Template}
\begin{figure*}[ht]
    \centering
    \includegraphics[width=0.9\linewidth]{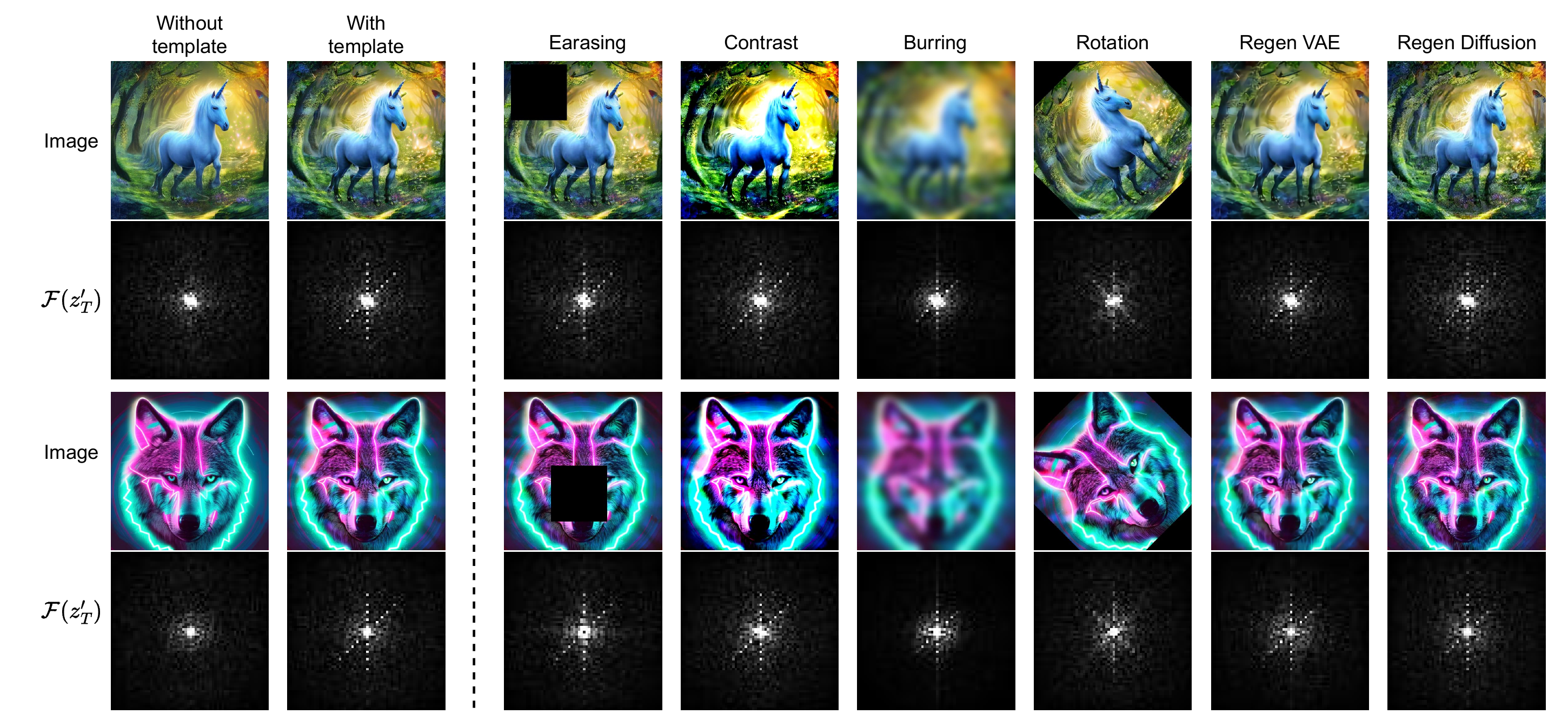}
    \caption{Visualization of the template space (second and fourth rows) under various distortions. To the left of the dashed line, images generated by Stable Diffusion 2.1 share the same prompt and initial noise, with template injections applied in the second column. The corresponding distorted template images are presented to the right of the dashed line.
    }
    \Description{\bfseries Figure descriptions should not repeat the figure caption – their purpose is to capture important information that is not already provided in the caption or the main text of the paper.}
    \label{template_result}
\end{figure*}
\label{analysis of template}
The template is desgined to be located in the middle-frequency region of the Fourier domain, where the image structure usually is located in the low-frequency region (\emph{i.e}., center of the $\mathcal{F}(z')$), while the details such as sharp edges are located in the high-frequency region (outer circle). Figure~\ref{template_result} visualizes the template in the Fourier domain under various distortions.

Actually, in the Fourier domain, the energy of each Fourier coefficient is influenced by all pixels in the original image. This characteristic makes the template resilient to erasing of deleting a portion of image content, as shown in Figure~\ref{template_result}—the template remains mostly intact. 
Instead of a complete loss of the template, the missing content impacts the lowest frequency components, which is reflected by the black dot in the center of the frequency spectrum.
The impact of other image manipulations on the template embedded in the middle frequencies of Fourier domain can be found in Figure~\ref{template_result} as well.
Basically, our extensive experiments demonstrate the robustness of the embedded templates.





\begin{figure}[h]
    \centering
    \includegraphics[width=0.95\linewidth]{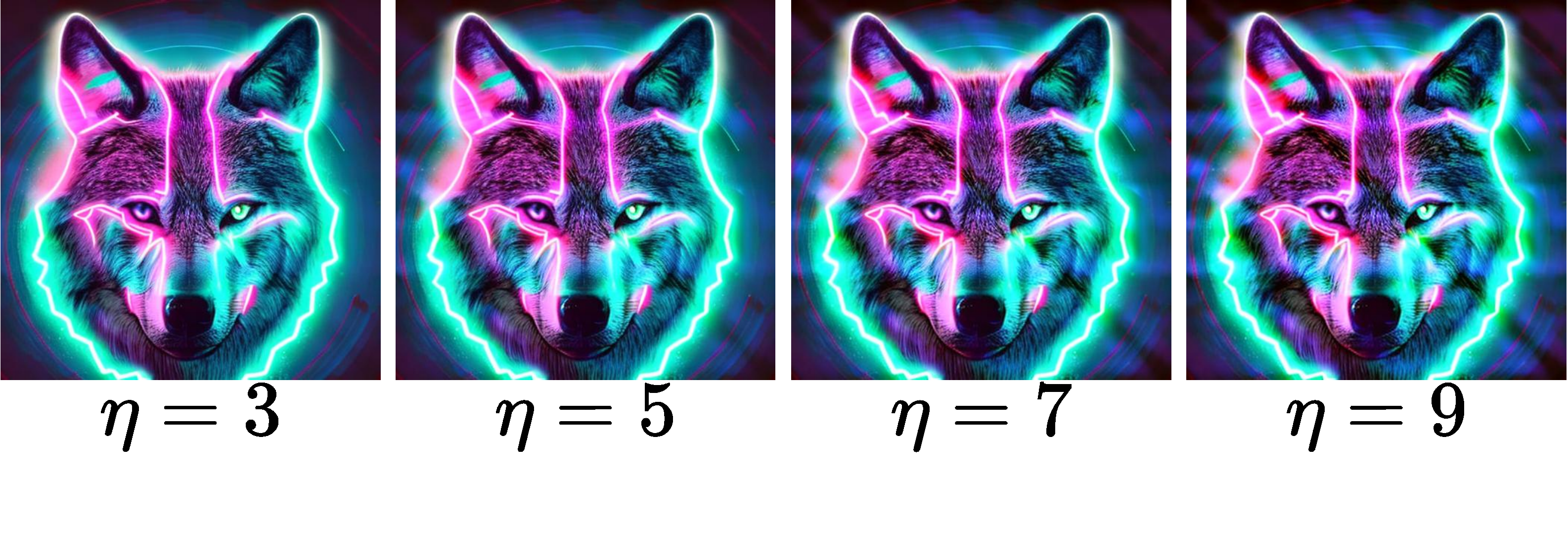}
    \caption{Visualization of different template strength.
    }
    \Description{\bfseries Figure descriptions should not repeat the figure caption – their purpose is to capture important information that is not already provided in the caption or the main text of the paper.}
    \label{ablation:template injection}
\end{figure}

\subsection{Tradeoff between Template Strength and Image Quality}
We investigate the effect of template strength on image quality through both visualizations (Figure~\ref{ablation:template injection}) and numerical results (Table \ref{abalation:template}). Table \ref{abalation:template} demonstrates that while stronger templates lead to a decrease in structural similarity, the FID remains in the same range, indicating that perceptual quality is preserved. The visualizations highlight only subtle changes in the image, yet our findings reveal that a stronger template tends to introduce a more pronounced pattern within the image space. This pattern formation suggests that while the impact on perceptual quality is minimal, the template strength can still influence the underlying structure of the image.


\begin{table}[ht]
    \caption{Template strength on image qualities.}\label{abalation:template}
    \centering
    \begin{tabular}{c|c  c c c}
    \toprule
        $\eta$ & SSIM & PSNR & FID \\
         \midrule
        1  & 0.78 & 21.00 &  17.10\\
        3  & 0.76 & 20.29 &17.32\\
        5  & 0.71  &19.07  & 17.35\\
        7  & 0.66 & 17.90&  17.85\\
        9  & 0.61  &16.69  & 18.60\\
        
        \bottomrule
    \end{tabular}
\end{table}

\section{Conclusion \& Limitations}


We present MaXsive, a high-capacity, robust, and training-free watermarking method for diffusion models. 
Unlike existing training-free watermarking methods that often trade capacity for robustness—resulting in vulnerabilities to RST attacks and potential ID collusion—MaXsive introduces an X-shaped watermarking template that significantly enhances robustness while preserving full watermark capacity. This innovative design enables MaXsive to achieve superior performance in both verification and identification scenarios, establishing it as a powerful and efficient training-free generative watermarking solution for real-world applications.

Resistance to cropping (accompanied with re-scaling in Stirmark) is still a challenge. 
As a future work, we will further investigate this issue.

\section{Acknowledgement}
This work was supported by the National Science and Technology Council (NSTC) with Grant NSTC 112-2221-E-001-011-MY2 and Academia Sinica with Grant AS-IAIA-114-M08. We thank to National Center for High-performance Computing (NCHC) for providing computational and storage resources.


\bibliographystyle{ACM-Reference-Format}
\bibliography{main}

\clearpage
\appendix
\clearpage

\section{Derivation of the RST Scaling Parameter}\label{sec:gamma}
In this section, we derive the formula for computing $\gamma$, which is used to scale an image back to its initial size before scaling.
Note that in the following discussion, we consider images of the same height and width.

Given a square image $\bm{I}$ of size $N\times N$, it is first rotated by $\theta$ degrees with its center (\emph{i.e}., $\left(\frac{N}{2},\frac{N}{2}\right)$) as the base point and then centrally cropped to produce an image $\bm{I}'$ of size $n\times n$ ($n \leq N)$, which is the largest possible image without any black padding.
Therefore, it is scaled to $N\times N$, making it feasible for diffusion models.
It is clear that the ratio of $n$ to $N$, denoted by $\gamma$, depends on $\theta$.
Thus, we derive $\gamma$ in terms of $\theta$.

Let the coordinates of $\bm{I}'$ be defined as
\begin{equation*}
    S = \left\{ (x,y) \left| \ \left| x-\frac{n}{2} \right|\leq\frac{n}{2},\ \left| y-\frac{n}{2} \right|\leq\frac{n}{2}\right\}\right.
\end{equation*}
and let $S$ be bounded by the region $R$, whose boundaries are given by the following four lines:
\begin{align}
    \left( x-\frac{n}{2} \right) \cos\theta &+ \left( y-\frac{n}{2} \right) \sin\theta = \frac{N}{2},\label{eq:L1} \\
    \left( x-\frac{n}{2} \right) \cos\theta &+ \left( y-\frac{n}{2} \right) \sin\theta = \frac{-N}{2},\nonumber \\
    -\left( x-\frac{n}{2} \right) \sin\theta &+ \left( y-\frac{n}{2} \right) \cos\theta = \frac{N}{2},\mbox{ and}\nonumber \\
    -\left( x-\frac{n}{2} \right) \sin\theta &+ \left( y-\frac{n}{2} \right) \cos\theta = \frac{-N}{2}\nonumber.
\end{align}
Since the rightmost boundary of $S$ is at $x=n$, we find the highest $y$ such that $(n,y)$ remains inside $R$, which is determined by Eq.~(\ref{eq:L1}).
Thus, 
\begin{equation*}
    \frac{n}{2}\cos\theta + (y-\frac{n}{2})\sin\theta = \frac{N}{2}
    \implies y = \left( \frac{N}{2\sin\theta} - \frac{n\cos\theta}{2\sin\theta} \right) + \frac{n}{2}.
\end{equation*}
Similarly, the top boundary of $S$ is at $y=n$; thus, for $S$ to be contained in $R$, we must have
\begin{equation*}
    n \leq \left( \frac{N}{2\sin\theta} - \frac{n\cos\theta}{2\sin\theta} \right) + \frac{n}{2},
\end{equation*}
which implies that
\begin{equation*}
    n \leq \frac{N}{\sin\theta + \cos\theta}.
\end{equation*}
Since a similar derivation holds when considering the constraints from the other boundaries, the largest possible $n$ is exactly $\frac{N}{\sin\theta + \cos\theta}$.
Hence, $\gamma$ is derived in terms of $\theta$ to be $\frac{1}{\sin\theta + \cos\theta}$.

\section{Exhaustive Evaluations on All Distortions in Attack Benchmarks}\label{Appendix: Details}

Actually, the Stirmark benchmark \cite{petitcolas1998attacks,petitcolas2000watermarking} has been widely adopted in the era of conventional non-learning-based watermarking community.
It contains extensive image manipulations, including both geometric distortions and non-geometric distortions.
The existing generative watermarking methods and learning-based post-processing watermarking methods are found to be insufficiently verified under such attacks.
In this subsection, we provide exhaustive verification results on Stirmark 3.1 in Table \ref{tab:stirmark_all}.

For new benchmark, WAVES \cite{an2024benchmarking}, both detailed verification and identification results are provided in Table \ref{tab:WAVES_all} and Table \ref{tab:WAVES_all_id}, respectively.

\begin{table*}[!t]
\centering
\caption{Verification on Stirmark 3.1}\label{tab:stirmark_all}
\resizebox{0.95\textwidth}{!}{
    \begin{tabular}{r|  c c c | c  }
    \toprule
    Distortions & Tree-Rings & RingID & Gaussian Shading& Our ($\eta = 5$)\\
    \midrule

     Median filter 2x2 & 0.53&1.00 &1.00 & 1.00\\
     Median filter 3x3 &0.48 &1.00 & 1.00&1.00\\
     Median filter 4x4 &0.33 &1.00 &1.00 &1.00\\
     Gaussian filter 3x3 &0.67 &1.00 &1.00 &1.00\\
     JPEG 90 &0.73 &1.00 &1.00 &1.00\\
     JPEG 80 &0.70 &1.00 &1.00 &1.00\\
     JPEG 70 &0.66 &1.00 &1.00 &1.00\\
     JPEG 60 &0.57 &1.00 &1.00 &1.00\\
     JPEG 50 &0.48 &1.00 &1.00 &1.00\\
     JPEG 40 &0.43 &1.00 &1.00 &1.00\\
     JPEG 35 & 0.42&1.00 &1.00 &1.00\\
     JPEG 30 & 0.40&1.00 &1.00 &1.00\\
     JPEG 25 &0.35 &1.00 & 1.00&1.00\\
     JPEG 20 &0.28 &1.00 & 1.00&1.00\\
     JPEG 15 & 0.20&1.00 & 1.00&1.00\\
     JPEG 10 &0.14 &1.00 & 1.00&0.99\\
     FMLR &0.36 &1.00 &1.00 &1.00\\
     Sharpening 3x3 &0.43 &1.00 &1.00 &1.00\\
     1 column, 1 row removed &0.68 &1.00 &1.00 &1.00\\
     5 column, 1 row removed & 0.54& 1.00&1.00 &1.00\\
     1 column, 5 row removed &0.59 &1.00 &1.00 &1.00\\
     17 column, 5 row removed &0.20 &1.00 &1.00 &1.00\\
     5 column, 17 row removed &0.42 & 1.00&1.00 &1.00\\
     Cropping 1\% off&0.27 &1.00 & 1.00&1.00\\
     Cropping 2\% off& 0.10&1.00 &1.00 &1.00\\
     Cropping 5\% off& 0.07&1.00 & 0.87&0.99\\
     Cropping 10\% off&0.01 &0.45 &0.10 &0.14\\
     Cropping 15\% off&0.01 &0.15 &0.02 &0.02\\
     Cropping 20\% off& 0.01&0.10 &0.03 &0.06\\
     Cropping 25\% off& 0.00&0.07 &0.01 &0.03\\
     Cropping 50\% off&0.00 &0.02 &0.01 &0.06\\
     Cropping 75\% off&0.00 &0.01 &0.01 &0.02\\
     Linear (1.007, 0.010, 0.010, 1.012) &0.15 & 1.00&1.00 &1.00\\
     Linear (1.010, 0.013, 0.009, 1.011) &0.13 &1.00 & 1.00&1.00\\
     Linear (1.013, 0.008, 0.011, 1.008) &0.14 &1.00 & 1.00&1.00\\

    \bottomrule
\end{tabular}
}
\end{table*}

\setcounter{table}{5}

\begin{table*}[!t]
\centering
\caption{Verification on Stirmark 3.1 (Cont.)}
\resizebox{0.95\textwidth}{!}{
    \begin{tabular}{r|  c c c | c  }
    \toprule
    Distortions & Tree-Rings & RingID & Gaussian Shading& Our ($\eta = 5$)\\
    \midrule

     Aspect ratio change (0.80, 1.00) &0.05 & 0.88& 0.87&0.48\\
     Aspect ratio change (0.90, 1.00) &0.19 & 1.00& 1.00&0.99\\
     Aspect ratio change (1.00, 0.80) & 0.03& 0.85& 0.72&0.60\\
     Aspect ratio change (1.00, 0.90) &0.11 &1.00 &1.00 &0.97\\
     Aspect ratio change (1.00, 1.20) &0.08 &0.96 & 0.96&0.48\\
     Aspect ratio change (1.00, 1.10) &0.21 &1.00 & 1.00&0.99\\
     Aspect ratio change (1.10, 1.00) &0.11 & 1.00& 0.99&0.99\\
     Aspect ratio change (1.20, 1.00) &0.11 & 0.90& 0.92&0.54\\
     Shearing x-0$\%$ y-1$\%$&0.19 & 1.00&1.00 &1.00\\
     Shearing x-1$\%$ y-0$\%$& 0.25&1.00 &1.00 &1.00\\
     Shearing x-1$\%$ y-1$\%$&0.17 &1.00 &1.00 &1.00\\
     Shearing x-0$\%$ y-5$\%$&0.07 &1.00 &0.89 &0.88\\
     Shearing x-5$\%$ y-0$\%$& 0.08&1.00 &0.85 &0.99\\
     Shearing x-5$\%$ y-5$\%$&0.06 &1.00 &1.00 &1.00\\
     Random bending&0.12 &1.00 &0.97 &1.00\\
     Rotation -2.00&0.10 &1.00 &0.93 &1.00\\
     Rotation -1.00&0.11 & 1.00&1.00 &1.00\\
     Rotation -0.75&0.10 &1.00 &1.00 &1.00\\
     Rotation -0.50&0.20 &1.00 &1.00 &1.00\\
     Rotation -0.25& 0.49 &1.00 &1.00 &1.00\\
     Rotation 0.25&0.50 & 1.00&1.00 &1.00\\
     Rotation 0.50& 0.22& 1.00&1.00 &1.00\\
     Rotation 0.75& 0.10& 1.00&1.00 &1.00\\
     Rotation 1.00&0.10 &1.00 &1.00 &1.00\\
     Rotation 2.00&0.10 &1.00 &0.95 &1.00\\
     Rotation 5.00&0.03 &0.72 &0.12 &0.59\\

    \bottomrule
\end{tabular}
}
\end{table*}

\setcounter{table}{5}
\begin{table*}[!t]
\centering
\caption{Verification on Stirmark 3.1 (Cont.)}
\resizebox{0.95\textwidth}{!}{
    \begin{tabular}{r|  c c c | c  }
    \toprule
    Distortions & Tree-Rings & RingID & Gaussian Shading& Our ($\eta = 5$)\\
    \midrule
     Rotation 10.00&0.01 & 0.16 &0.02 &0.66\\
     Rotation 15.00&0.01 &0.07 &0.01 &0.63\\
     Rotation 30.00&0.00 & 0.04&0.01 &0.66\\
     Rotation 45.00&0.00 &0.02 &0.00 &0.62\\
     Rotation 90.00&0.02 &0.99 &0.01 &1.00\\
     Rotation scale -2.00&0.11 &1.00 &0.91 &1.00\\
     Rotation scale -1.00& 0.19&1.00 &1.00 &1.00\\
     Rotation scale -0.75& 0.10 & 1.00& 1.00&1.00\\
     Rotation scale -0.50&0.16 & 1.00&1.00 &1.00\\
     Rotation scale -0.25& 0.47 & 1.00& 1.00&1.00\\
     Rotation scale 0.25& 0.48 &1.00 & 1.00&1.00\\
     Rotation scale 0.50& 0.17 &1.00 & 1.00&1.00\\
     Rotation scale 0.75& 0.10 & 1.00& 1.00&1.00\\
     Rotation scale 1.00& 0.19 &1.00 &1.00 &1.00\\
     Rotation scale 2.00& 0.09 &1.00 &0.95 &1.00\\
     Rotation scale 5.00& 0.03 &0.72 &0.10 &0.56\\
     Rotation scale 10.00&0.01 & 0.19& 0.03&0.66\\
     Rotation scale 15.00&0.01 &0.11 &0.01 &0.64\\
     Rotation scale 30.00&0.00 &0.03 &0.01 &0.67\\
     Rotation scale 45.00&0.01 &0.02 &0.04 &0.64\\
     Rotation scale 90.00& 0.02 &0.99 &0.01 &1.00\\
     scale 2.00&0.71 &1.00 &1.00 &1.00\\
     scale 1.50&0.73 & 1.00&1.00 &1.00\\
     scale 1.10&0.65 &1.00 &1.00 &1.00\\
     scale 0.90&0.66 & 1.00&1.00 &1.00\\
     scale 0.75&0.66 &1.00 &1.00 &1.00\\
     scale 0.50&0.48 & 1.00&1.00 &1.00\\
    \bottomrule
\end{tabular}
}
\end{table*}

\begin{table*}[t]
\centering
\caption{Verification on WAVES}\label{tab:WAVES_all}
\resizebox{0.95\textwidth}{!}{
    \begin{tabular}{r|  c c c  |c c |c c c |c c  | c c  }
    \toprule
    &   \multicolumn{3}{c}{ Geometric}&\multicolumn{2}{|c}{Photometric}&\multicolumn{3}{|c|}{Degradation}&\multicolumn{2}{c|}{Regeneration}&\multicolumn{2}{|c}{Adversarial}\\
    Method  & Rotation & C\&R &Erasing  &  Contrast & Brightness & Blurring &  Noise & JPEG & VAE & Diff& CLIP & ResNet \\
    \midrule
    Tree-Rings& 0.01 & 0.02 & 0.60 & 0.53& 0.48  &  0.01 & 0.33 & 0.43 & 0.07 & 0.07 & 0.40 & 0.37 \\
    RingID & 0.99 & 0.15 &\textbf{1.00} &  \textbf{1.00}& \textbf{1.00}  &  0.87 & \textbf{1.00} & \textbf{1.00} &  \textbf{1.00} &  \textbf{1.00} &  \textbf{1.00} &\textbf{1.00}\\
    Gaussian Shading &  0.26 & \textbf{0.47}  & \textbf{1.00} &\textbf{1.00}& \textbf{1.00}&  \textbf{0.92} &\textbf{1.00} &\textbf{1.00} &\textbf{1.00}&\textbf{1.00}&  \textbf{1.00}&\textbf{1.00}\\
    \midrule
    
    Our ($\eta = 5$)&\textbf{1.00} & 0.20 & \textbf{1.00}  & \textbf{1.00} & \textbf{1.00}  &0.85&\textbf{1.00} &\textbf{1.00}& \textbf{1.00} & \textbf{1.00} & \textbf{1.00}& \textbf{1.00}  \\
    \bottomrule
\end{tabular}
}
\end{table*}

\begin{table*}[!t]
\centering
\caption{Identification on WAVES}\label{tab:WAVES_all_id}
\resizebox{0.95\textwidth}{!}{
    \begin{tabular}{r|  c c c  |c c |c c c |c c | c c }
    \toprule
    &  \multicolumn{3}{c}{ Geometric}&\multicolumn{2}{|c}{Photometric}&\multicolumn{3}{|c|}{Degradation}&\multicolumn{2}{c}{Regeneration}&\multicolumn{2}{|c}{Adversarial}\\
    Method &  Rotation & C\&R &Erasing  &  Contrast & Brightness & Blurring &  Noise & JPEG & regen & adv &CLIP &ResNet  \\
    \midrule
    Tree-Rings&  0.00 & 0.00 & 0.04 & 0.05 & 0.03 & 0.00 &0.01 & 0.01&0.01 & 0.01&0.04 & 0.04  \\
    RingID &  0.34 & 0.05 &  0.42 & 0.42 &0.41& 0.17 & 0.41 & 0.40 & 0.40 & 0.40& 0.42 & 0.42  \\
    Gaussian Shading&  0.00 & 0.10 & \textbf{1.00} & \textbf{1.00} & \textbf{1.00} & 0.18 & \textbf{1.00} & 0.99 & 0.99& 0.99& \textbf{1.00} & \textbf{1.00}  \\\midrule
    Our ($\eta = 5$)&\textbf{1.00} & \textbf{0.48} & \textbf{1.00} & \textbf{1.00} & \textbf{1.00} & \textbf{0.79} & \textbf{1.00} & \textbf{1.00}& \textbf{1.00} & \textbf{1.00}& \textbf{1.00} & \textbf{1.00}  \\
    \bottomrule
\end{tabular}
}
\end{table*}

\end{document}